\documentstyle[twoside,fleqn,espcrc2,subfloat,epsbox]{article}
\input epsf
\title{\bf Confinement Properties in the Multi-Instanton System}
\author{\underline{M. Fukushima}
\address{Research Center for Nuclear Physics (RCNP), Osaka University, 
Ibaraki, Osaka 567, Japan}\thanks{ e-mail: masa@rcnp.osaka-u.ac.jp},
H. Suganuma $^{\rm a}$, A. Tanaka $^{\rm a}$, H. Toki $^{\rm a}$ and S. Sasaki
\address{Yukawa Institute for Theoretical Physics, Kyoto University,
Kyoto 606-01, Japan}
}

\begin{document}

% abstract
\begin{abstract}
\vspace{0.3cm}
We investigate the confinement properties in the multi-instanton system, 
where the size distribution is assumed to be $ \rho^{-5} $ for the large 
instanton size $ \rho $. 
We find that the instanton vacuum gives the area law behavior of the Wilson loop,
which indicates existence of the linear confining potential. 
In the multi-instanton system, the string tension increases monotonously 
with the instanton density, and takes the standard value $ \sigma \simeq 1 
{\rm GeV/fm} $ for the density $ (N/V)^{\frac{1}{4}} = 200 {\rm MeV} $.
Thus, instantons directly relate to color confinement properties.\\
\noindent

\vspace{0.0cm}
\end{abstract}

\maketitle

% Confinement Properties and Topological Objects in QCD vacuum
\section{Topological Objects in QCD Vacuum}
In the QCD vacuum, there are two non-trivial topological objects, instantons 
and monopoles, which belong different sectors of physics. It is believed that these objects are 
independently important for understanding the non-perturbative properties.
An instanton appears as a classical and non-trivial solution in the Yang-Mills 
theory corresponding to the homotopy group, $\pi_{3}(SU(N_{c}))=Z_{\infty}$ 
\cite{polya }.
Instantons are important for the phenomena related to the $U_{A}(1)$
anomaly and chiral symmetry breaking \cite{Diak ,Shur }.
%\noindent 
%\vspace{-0.3cm}
On the other hand, QCD is reduced to an abelian gauge theory 
with QCD-monopoles after performing 
the abelian gauge fixing \cite{hooft }. QCD-monopoles appear as 
the topological defects corresponding to the nontrivial homotopy group
$\pi_{2}(SU(N_{c})/U(1)^{N_{c}-1})=Z_{\infty}^{N_{c}-1}$.
Condensation of monopoles plays an essential role on color confinement 
and chiral symmetry breaking \cite{suga_1}-\cite{miya_2}. \\
\noindent 

\vspace{-0.4cm}
Until now, there has been no evidence that the instanton has anything to 
do with color confinement. 
However, recent studies \cite{miya_2}-\cite{fuku_2} suggest that instantons 
directly relate to monopoles, 
whose condensation provides an interpretation of the confinement mechanism.
In this paper, we study the further relation between instantons
and the confinement properties. \\
\noindent 

\vspace{-0.5cm}
% monopole condensation in the multi-instanton vacuum 
\section{Correlation between \\
\indent \hspace{1.0cm} Instantons and Monopoles}
Recently, it has been found that there exists a strong correlation between 
instantons and monopoles in the abelian projected theory of QCD, 
both in the analytical and lattice frameworks \cite{miya_2}-\cite{suga_4}. 
In the maximally abelian gauge, a monopole trajectory seems to be localized 
around the center of an instanton \cite{Hart ,Brow }. We studied the multi-instanton 
system in terms of monopole condensation numerically \cite{fuku_1,fuku_2}. 
When the instanton density is high and instantons overlap each other, 
there appears an very long and highly complicated monopole trajectory.
This monopole trajectory covers the entire physical volume ${\bf R}^4$ in 
the multi-instanton system.
\cite{fuku_1,fuku_2}.
In the lattice QCD simulation, the large monopole clustering is observed 
in the confinement phase, which includes many instantons and anti-instantons as 
the topological excitation \cite{suga_4}. Thus,
such long and complicated monopole trajectory is a signal of monopole condensation, 
which is responsible for color confinement.\\
\noindent 

\vspace{-0.4cm}
Therefore, instantons would play a relevant role on the confinement 
mechanism via the promotion of monopole condensation.
In order to clarify the relation of instantons with color confinement, we study 
the multi-instanton system, which is regarded to hold the essence of the 
non-perturbative QCD vacuum.\\
\noindent 

\vspace{-0.9cm}
% Multi-Instanton system
\section{Multi-Instanton Configuration}
The gauge configuration of an instanton with the size $\rho$ and the center 
$z$ in the singular gauge is expressed as \\
\noindent 

\vspace{-0.5cm}
\begin{equation}
A_{\mu }^{I}(x;z,\rho,O)=
\frac {i\tau^{a} \rho^{2}O^{ab} \bar{\eta }^{b}_{\mu \nu}(x-z)_{\nu}}
{(x-z)^{2}[(x-z)^{2}+\rho^{2}]},
\end{equation}
\noindent 

\vspace{-0.1cm} 
\noindent 
where $O^{ab}$ denotes the SU(2) color orientation matrix and 
$\bar{\eta }^{a}_{\mu \nu } $ the 't Hooft symbol. 
The anti-instantons configuration $A_{\mu }^{\bar{I}}$ is obtained 
by replacing $\bar{\eta } ^{a}_{\mu \nu }$ with $\eta ^{a}_{\mu \nu } $. \\
\noindent 

\vspace{-0.4cm}
The multi-instanton ensemble is characterized by the size distribution,
the randomness of the color orientation and the density of instantons \cite{fuku_1,fuku_2}. 
The size distribution has to follow the one loop result,
$f(\rho )\sim \rho^{b-5}$ with $b=11N_{c}/3$, in the ultra-violet region. 
For the large size, the ordinary instanton liquid models suggest that 
the size distribution behaves as $f(\rho )\sim 1/\rho^{5}$ due to the 
infrared repulsive force \cite{Shur }. Therefore, we assume the size distribution as\\
\noindent 

\vspace{-0.5cm}
\begin{equation}
f(\rho )=\frac{1}{(\frac{\rho}{\rho_{1}})^{\nu }+(\frac{\rho_{2}}{\rho})^{b-5} }
\label{eqn:distri}
\end{equation}
\noindent 

\vspace{-0.2cm} 
\noindent 
with size parameters $\rho_{1}$ and $\rho_{2}$, which should satisfy the 
normalization condition, $\int^{\infty }_{0} d\rho f(\rho )=1$. The maximum 
of the distribution is fixed to the standard probable size $\rho_{0}=0.4{\rm fm}$.
In this calculation, we take $\nu = 5$ for the behavior of the size distribution
on large instantons, although we are planning to investigate how 
the long-range physics depend on the infrared behavior of the size distribution.\\ 
\noindent 

\vspace{-0.4cm}
The multi-instanton configurations are approximated as the sum of instanton 
and anti-instanton solutions, \\
\noindent
 
\vspace{-0.9cm}
\begin{eqnarray}
A_{\mu }(x)=\sum_{k} A_{\mu }^{I}(x;z_{k},\rho_{k},O_{k}) \nonumber \\
\vspace{-0.5cm} & & \makebox[-2.0cm]{} 
+\sum_{\bar{k}} A_{\mu }^{\bar{I}}(x;z_{\bar{k}},\rho_{\bar{k}},O_{\bar{k}}).
\end{eqnarray}   
\noindent 

\vspace{-0.2cm} 
\noindent 
We generate the ensemble of pseudoparticles with random color orientations $O_{k}$ 
and random centers $z_{k}$. The instanton sizes $\rho_{k}$ are randomly taken 
according to Eq.(\ref{eqn:distri}). In spite of simple sum ansatz for 
multi-instanton system, {\it the interaction among these pseudoparticles 
would be included effectively in the instanton size distribution}.\\
\noindent 

\vspace{-0.4cm}
Above procedures are performed in the continuum theory. 
Here, we introduce a lattice on this gauge configuration and define the 
link variables, $U_{\mu }(s)={\rm exp}(iaA_{\mu }(s))$. 
The string tension is estimated from the SU(2) Wilson loop,\\
\noindent 

\vspace{-0.4cm}
\begin{equation}
W[C] \equiv {\rm tr} P \exp [ i\oint_{C}A_{\mu}dx_{\mu}]
={\rm tr}{\prod_{C}U_{\mu}(s)},
\end{equation}
\noindent 

\vspace{-0.2cm} 
\noindent 
also in the multi-instanton system. In order to smooth out the ultra-violet noise, 
we will use the smeared link variable,
\noindent 

\vspace{-0.1cm}
\begin{equation}
\tilde{U}_{\mu} \equiv \frac{
U_{\mu} \! + \! \! \displaystyle{\sum_{\nu \ne \pm \mu }}
U_{\nu}(s) U_{\mu}(s+\hat{\nu}) U_{\nu}^{\dagger}(s+\hat{\mu})
}{ \parallel
U_{\mu} \! + \! \! \displaystyle{\sum_{\nu \ne \pm \mu }}
U_{\nu}(s) U_{\mu}(s+\hat{\nu}) U_{\nu}^{\dagger}(s+\hat{\mu})
\parallel },
\end{equation}
\noindent 

\vspace{-0.2cm} 
\noindent 
instead of $U_{\mu}$ in the actual calculation, although such replacement 
never changes the string tension expect for reduction of the error bar.\\
\noindent 

\vspace{-0.5cm}
% Our Numerical Results
\section{Results and Concluding Remarks}
We take the standard instanton density 
$(N/V)^{\frac{1}{4}} = 1 {\rm fm}^{-1} = 200 {\rm MeV}$ \cite{Diak ,Shur }
with the equal numbers of instantons and anti-instantons; 
$N_{I} = N_{\bar{I}} = N/2 $.
We adopt the lattice with the total volume $V=(3.0 {\rm fm})^{4}$,
and take the different lattice spacing $a=0.150 {\rm fm}$ and 
$a=0.125 {\rm fm}$ on the $20^{4}$ and $24^{4}$ lattices, respectively.\\
\noindent 

\vspace{-0.4cm}
Fig.1 shows the Wilson loop behavior in the multi-instanton system.
In the long-range region, the Wilson loop obeys the area law, which indicates 
existence of the linear confining potential. 
As shown in Fig.2, the static quark potential $V(R)$ increases in the 
infrared region, and is approximately proportional to the inter-quark 
distance $R$. It is remarkable that the string tension is obtained as 
$ \sigma \simeq 1 {\rm GeV}/{\rm fm} $ in this multi-instanton system,
which agrees with the standard value.\\
\noindent 

\vspace{-0.4cm}
By comparing the two different lattice spacing cases, above results are 
expected to hold in continuum limit. Further analyses show that  
{\it small instantons do not contribute the linear confining potential and 
large instantons play an essential role on the long-range physics}.\\
%At long-range region, there clearly appears an area law behavior, which indicates the 
%existence of a linear confining potential. Compared with two different lattice 
%spacing cases, such area law behavior seems to survive continuum limit.
%As shown in Fig.2, the potential $V(R)$ tends to increase at the infrared region 
%and is approximately proportional to the distance $R$ between a charged pair. 
%The string tension $ \sigma $ of this system is close to
%$ 1 {\rm GeV}/{\rm fm}$, which is the value corresponding to the original QCD vacuum.
%We find that small instantons do not contribute the linear
%confining potential and large instantons play an essential role on the long range 
%physics.\\
\noindent 

\begin{figure*}[hbt]
\hspace{-0.3cm}
\begin{minipage}[hbt]{5cm}
\epsfxsize = 5cm
\epsfbox{Fig_01.EPSF}
\end{minipage}
\hspace{0.2cm}
\begin{minipage}[hbt]{5cm}
\epsfxsize = 5cm
\epsfbox{Fig_02.EPSF}
\end{minipage}
\hspace{0.2cm}
\begin{minipage}[hbt]{5cm}
\epsfxsize = 5cm
\epsfbox{Fig_03.EPSF}
\end{minipage}
\noindent
\vspace{0.0cm}\\

\hspace{0.0 cm}
\begin{minipage}[hbt]{4.5cm}
Fig.1: The Wilson loop $\ln W[R,T]$ vs.$\!$ area $R\!\times\! T$ in
the multi-instanton system. 
\end{minipage}
\hspace{0.7cm}
\begin{minipage}[hbt]{4.5cm}
Fig.2: The static quark potential in the multi-instanton system. 
\end{minipage}
\hspace{0.7cm}
\begin{minipage}[hbt]{4.5cm}
Fig.3: The static quark potential for various instanton densities.
\end{minipage}
\vspace{0.15cm}
\end{figure*}

\vspace{-0.4cm}
Finally, Fig.3 shows the static potential behaviors for various instanton densities.
We take the four cases with $(N/V)^{\frac{1}{4}} =125$, $150$, $175$ and 
$200 {\rm MeV}$, and use the $20^{4}$ lattice with $a=0.150 {\rm fm}$.
As the instanton density becomes lower, the slope of the potential tends to 
decrease monotonously. This tendency indicates that the string tension 
depends on the instanton density directly. 
Therefore, the disappearance of instantons near the critical temperature 
would be important for the deconfinement phase transition.  \\
\noindent

\vspace{-0.4cm}
In conclusion, the main feature of the non-perturbative QCD vacuum would be
characterized by instantons and anti-instantons, which cause the color 
randomness and provide monopole condensation \cite{fuku_1,fuku_2}. 
We have studied that the confinement properties in the multi-instanton 
system by using the numerical analyses of the Wilson loop.
In the infrared region, the static quark potential is found to be 
proportional to the inter-quark distance, $V(R) \simeq \sigma R$.
For the standard instanton density  $(N/V)^{\frac{1}{4}} = 200 {\rm MeV} $,
we have obtained $ \sigma \simeq 1 {\rm GeV}/{\rm fm} $ in the 
multi-instanton system. Our numerical study suggests that instantons play 
an substantial role not only monopole condensation but also the 
confinement properties directly.
%Thus, we investigate the Wilson loop behavior of the multi-instanton systems with 
%various instanton densities numerically. We find that the potential is 
%approximately proportional to the distance, $ V(R) \sim \sigma R $. 
%In the instanton density $(N/V)^{\frac{1}{4}} = 200 {\rm MeV} $, the string tension 
%$ \sigma \sim 1 {\rm GeV}/{\rm fm} $ is obtained.
%Our numerical investigation shows that instantons may play an substantial role not only 
%on monopole condensation, but also the confinement properties.\\
\noindent 

\vspace{0.1cm}
% bibliography

%\begin{thebibliography}{9}
%\bibitem{Polya } A.~Belavin, A.~Polyakov, A.~Shvarts and Yu.~Tyupkin,~Phys.~Lett.~{\bf 59B}~(1975)~85.
%\bibitem{Hooft } G.~'t~Hooft,~Nucl.~Phys.~{\bf B190}~(1981)~455.
%\bibitem{suga_1} H.~Suganuma, S.~Sasaki and H.~Toki,~Nucl.~Phys.~{\bf B435}~(1995)~207.
%\bibitem{miya_1} O.~Miyamura,~Nucl.~Phys.~{\bf B}~(Proc.Suppl.)~{\bf 42}~(1995)~538.
%\bibitem{miya_2} O.~Miyamura~and S.~Origuchi~in~{\it Color~Confinement~and~Hadrons}~(World~Scientific,1995)~235.
%\bibitem{suga_2} H.~Suganuma,~H.~Ichie,~S.~Sasaki~and~H.~Toki,~in~{\it Color~Confinement~and~Hadrons}~(World~Scientific,1995)~65.
%\bibitem{markum} S.~Thurner,~H.~Markum~and~W.~Sakuler,~in~{\it Color~Confinement~and~Hadrons}~(World~Scientific,1995)~77.
%\bibitem{suga_3} H.~Suganuma,~A.~Tanaka,~S.~Sasaki~and~O.~Miyamura,~Nucl.~Phys.~{\bf B}~(Proc.Suppl.)~{\bf 47}~(1996)~302.
%\bibitem{Cher  } M.~Chernodub~and~F.~Gubarev,~JETP~Lett.{\bf 62}~(1995)~100.
%\bibitem{Hart  } A.~Hart~and~M.~Teper,~Phys.~Lett.~{\bf B371}~(1996)~261.
%\bibitem{Brow  } R.~Brower~K.~Originos~and~C.~Tan,~Phys.~Rev.~{\bf D55}~(1997)~6313.
%\bibitem{suga_4} H.~Suganuma,~S.~Sasaki,~H.~Ichie,~H.~Toki~and~F.~Araki,~in~{\it Frontier '96}~(World Scientific).
%\bibitem{fuku_1} M.~Fukushima,~A.~Tanaka,~S.~Sasaki,~H.~Suganuma,~H.~Toki~and~D.~Diakonov,~Nucl.~Phys.~{\bf B}~(Proc.Suppl.)~{\bf 53}~(1997)~494. 
%\bibitem{fuku_2} M.~Fukushima,~S.~Sasaki,~H.~Suganuma,~A.~Tanaka,~H.~Toki~and~D.~Diakonov,~Phys.~Lett.~{\bf B399}~(1997)~141.
%\end{thebibliography}

\end{document}